\documentclass[aps,pre,a4paper,twocolumn,amsmath,showpacs,floatfix]{revtex4}
\usepackage{graphicx}

\begin{document}
\bibliographystyle{apsrev}

\title{Bulk inhomogeneous phases of anisotropic particles: 
A fundamental measure 
functional study of the restricted orientations model}

\author{Yuri Mart\'{\i}nez-Rat\'on}
\email{yuri@math.uc3m.es}
\affiliation{Grupo Interdisciplinar de Sistemas Complejos (GISC), \\ 
Departamento de Matem\'aticas, Universidad Carlos III de Madrid, \\
Avenida de la Universidad 30, E--28911, Legan\'es, Madrid, Spain.} 

\date{\today}

\begin{abstract}
The phase diagram of prolate and oblate particles in 
the restricted orientations 
approximation (Zwanzig model) is calculated. Transitions 
to different inhomogeneous phases 
(smectic, columnar, oriented, or plastic 
solid) are studied through minimization 
of the fundamental measure functional (FMF) of hard parallelepipeds. The 
study of parallel hard cubes (PHC's) as a particular case 
is also included motivated 
by recent simulations of this system. As a result 
a rich phase behavior is obtained which include, apart from the usual liquid 
crystal phases, a very peculiar phase (called here discotic smectic) 
which was already found in the only existing simulation of the model, 
and which turns out to be stable because of the restrictions 
imposed on the orientations. The 
phase diagram is compared at a qualitative level with  
simulation results of other anisotropic particle systems.  

\end{abstract}

\pacs{64.70.Md,64.75.+g,61.20.Gy}
% 64.70.Md  Transitions in liquid crystals
% 64.75.+g  Solubility, segregation, and mixing; phase separation
% 61.20.Gy  Theory and models of liquid structure   

\maketitle

\section{Introduction}
\label{Introduccion}

Onsager first showed that the isotropic-nematic 
liquid crystal phase transition occurs  
in systems of anisotropic particles interacting via  
hard core repulsions \cite{Onsager}. 
He studied a system of hard spherocylinders in 
the limit of infinite anisometry $\kappa=(L+D)/D\to \infty$ 
($\kappa$ is the spherocylinder length to breath ratio) 
using the second virial 
form of the free energy, which in this limit is exact for the isotropic 
phase. The effect that  
higher virial coefficients have in the isotropic-nematic transition was later 
studied by Zwanzig, who introduced a model of hard prolate uniaxial 
parallelepipeds with axes oriented along the three perpendicular 
directions \cite{Zwanzig}.
This peculiar model, which obviously treats 
the orientational degrees of freedom 
in an unrealistic way, has the advantage of being accessible to the calculation 
of higher virial coefficients up to seventh order in the infinite aspect ratio 
limit. He showed that including 
all these virials the isotropic-nematic transition also occur, although 
the exact value of the coexisting nematic density strongly depends on 
the order of the approximation.  The Pad\'e approximant generated by the 
truncated cluster expansion provides a much more stable sequence of the 
parameters which characterize the transition \cite{Runnels}. This stability 
leaves little room for doubts regarding the existence of the transition in 
the model. The virial expansion resumed and expressed in the variable 
$y=\rho/(1-\rho v)$, with $\rho$ the number density 
of parallelepipeds and $v$ their volume, 
converges more rapidly than the traditional expansion in $\rho$, as was 
shown by Barboy and Gelbart for different hard particle geometries 
\cite{Barboy}. Thus, the so-called $y_3$ expansion of the Zwanzig model 
was applied to the study of the isotropic-nematic transition as well as  
to the study of properties of its interface \cite{Moore}. 
For the latter the authors applied 
the smoothed density approximation of the free energy functional
in the spirit of Tarazona's weighted density approximation 
for the fluid of hard spheres \cite{Tarazona}. 

The restricted orientations model for hard cylinders was 
also used to describe the 
structural properties of molecular fluids near hard walls or confined in a 
slit. This time the density functional was constructed from the bulk 
direct correlation function approximated by a linear combination of 
geometrical functions \cite{Rickayzen}.

Computer simulations of a variety of models of nonspherical hard core 
particles showed that the excluded volume 
effects could not only account 
for the stability of nematics, but also for the existence of liquid 
crystal inhomogeneous phases such as the smectic \cite{Strobants_1}
and columnar \cite{Veerman} phases. Particularly the 
complete phase diagram of freely rotating 
hard spherocylinders \cite{Bolhuis}, including not only 
smectic, but also a plastic solid phase and different 
oriented solid phases was calculated. Several 
density functional theories, all of them based on weighted or modified 
weighted density approximations, are able to reproduce reasonably well 
the isotropic-smectic or nematic-smectic transitions 
\cite{Somoza,Poniewierski,Velasco} 
in the whole range of aspect ratios where the smectic is stable, and in some 
cases,  transitions from the isotropic fluid to the plastic 
or oriented solid phases \cite{Graf}. In all these 
approximations the excess free energy is evaluated by integration of 
the free energy per particle of a reference fluid 
(typically spheres or hard parallel ellipsoids) 
evaluated at some weighted or effective 
density. In some cases, the employed weight is directly the normalized Mayer 
function between spherocylinders \cite{Poniewierski,Velasco}; 
 in others, it is calculated from the knowledge 
of the bulk correlation function of the reference fluid \cite{Velasco}. 
For the latter case, 
the term proportional to the Mayer function enters into the integrand 
as a multiplicative factor of the free energy per particle. 
The hard sphere free energy 
functional is recovered in both approaches as the limiting case of $L=0$. 

The fundamental measure theory (FMT) first developed for hard spheres by 
Rosenfeld \cite{Rosenfeld_1} was another starting point for constructing 
a density functional for anisotropic particles. In its general 
formalism the excess free energy density of the fluid is a function 
of some weighted densities obtained by convoluting the density profiles with  
weights which are characteristic functions of the geometry of a single 
particle whose integrals are the so-called fundamental measures: 
volume, surface area, and mean radius of the particles. Unfortunately, 
the Mayer function of two convex anisotropic bodies 
cannot be decomposed as a finite sum of convolutions of single particle 
weights \cite{Rosenfeld_2}, which is the keystone 
for constructing such a functional. Thus, the 
low density limit of the direct correlation function is no more the 
Mayer function. 

In spite of this, Chamoux and Perera have taken advantage of 
Rosenfeld's extension of FMT to hard convex bodies by using it to compute 
the direct correlation function and patching out the low density limit with 
the exact Mayer function \cite{Chamoux_1}. In this way they have obtained 
the equation of state for various convex hard bodies (such as hard ellipsoids, 
spherocylinders, and cut spheres), have predicted ordered phases and, 
recently, have study demixing in binary mixtures of rigidly ordered 
particles \cite{Chamoux_1,Chamoux_2}.  

Following a similar procedure a density functional for   
anisotropic particles has been proposed which 
interpolates between the Rosenfeld's hard sphere  
functional and Onsager's functional for elongated rods. The 
resulting model was tested by calculating the isotropic-nematic transition 
in systems of hard spherocylinders and hard ellipsoids \cite{Cinacchi}.

Although the authors of this work suggest that the 
resulting theory can be applied 
to the study of inhomogeneous systems, the huge computational efforts that 
their numerical implementations involve is the reason for the absence of any 
result in this direction. 
One way to circumvent this difficulty is to 
reduce the continuous orientational degree of freedom to three  
discrete orientations (Zwanzig model). Implementing this idea  some authors 
have recently applied 
the Zwanzig model to the study of interfacial properties of the hard 
rod fluid interacting with a hard wall or confined in a slit, 
for a one-component 
\cite{vanRoij} and a polydisperse mixture \cite{Martinez-Raton_1}, 
and also to the study 
of bulk and interfacial properties of  hard platelet binary mixtures 
\cite{Harnau}. All these models are based on Onsager's density 
functional approximation. The increase of the number of allowed orientations 
in this functional particularized for hard 
spherocylinders results in the presence 
of an artificial nematic-nematic transition in the one component fluid as 
the authors of Ref. \cite{vanRoij_1} have shown. This result indicates 
that certain cares must be taken in the direct extrapolation of the results 
obtained from this theory.   

FMF was 
also constructed for a mixture of parallel hard cubes combining 
Rosenfeld's original ideas with a  
dimensional cross over constraint \cite{Cuesta}. 
The latter appears to be very important to describe correctly 
the structure of inhomogeneous fluids in situations of high confinement 
and to describe well the structural properties of 
the solid phase \cite{Tarazona_1}. The dimensional cross over has been used 
as an important ingredient to develop a density functional for a binary 
mixture of hard spheres and needles, assuming that the needles are too 
thin to interact with each other directly \cite{Schmidt}. 

Taking a ternary mixture of parallel hard 
cubes and scaling each species along one of the three 
Cartesian axes with the same scaling 
factor a FMF for the Zwanzig model is obtained. This functional has 
already been applied to the study of the effect that polydispersity has on the
stability of the biaxial phase in a binary mixture of rods and plates 
\cite{Martinez-Raton_2} and on the relative stability of 
the smectic and columnar phases
due to the presence of polydispersity \cite{Martinez-Raton_3}. 

The FMF for Zwanzig's model in the homogeneous limit coincide with the scaled 
particle theory and thus with the so-called 
$y_3$ expansion which, as pointed 
out before, first began to be used in Ref. \cite{Moore} as a model to study 
the isotropic-nematic 
phase transitions in fluids of hard parallelepipeds. But this 
functional, through its minimization, also allows us to calculate  
inhomogeneous density profiles. 
This functional has been applied recently to study the isotropic-nematic 
interface of a binary mixture of hard platelets \cite{Bier}. 
Its structural and thermodynamic 
properties resulting from the FMF minimization
show complete wetting by a second nematic. The same phenomenon  
was found in a binary mixture of hard spherocylinders \cite{Shundyak}.  

The phase diagram for Zwanzig's model
including the smectic, columnar, and 
solid phases has never been carried out, only spinodal instability 
boundaries have been traced \cite{Martinez-Raton_3}. 
The main purpose of this work is to obtain the complete 
phase diagram for this model and to compare the results with the only 
existing simulation of the lattice version of the model, which has been 
carried out for two different aspect ratios \cite{Casey}. This will test 
the predictive power of the FMF for anisotropic inhomogeneous phases. 
As a particular case, the system of parallel 
hard cubes will be studied. In Ref. \cite{Martinez-Raton_4} 
a bifurcation analysis 
and a Gaussian parametrization of the density profiles were used to calculate
the free energy and pressure of the solid phase. Here 
a free minimization will be performed to calculate 
not only the solid but also the smectic and columnar phases and compare 
the obtained results with recent simulations of parallel hard cubes 
\cite{Jagla,Groh}.

\section{FMF for Zwanzig model}
\label{FMF}
The FMF for hard parallelepipeds was 
already described in detail in Ref. \cite{Cuesta}.  
A brief summary of the theory will be presented here 
putting emphasis on its numerical implementation to calculate the 
equilibrium inhomogeneous phases.

A ternary mixture of hard parallelepipeds of cross section $\sigma^2$ 
and length $L$ with their uniaxial axes pointing 
to the $x$, $y$, or $z$ directions is described in terms of 
their density profiles $\rho_{\mu}({\bf r})$ ($\mu=x,y,z$).  
Following the FMT for hard parallelepipeds in 
three dimensions the excess free energy density in reduced units 
can be written as \cite{Cuesta} 
\begin{eqnarray}
\Phi_{\rm{exc}}({\bf r})=\Phi^{(1)}({\bf r})+\Phi^{(2)}({\bf r})+
\Phi^{(3)}({\bf r}), \label{three}
\end{eqnarray}
where the $\Phi^{(k)}$'s are
\begin{eqnarray}
\Phi^{(1)}&=&-n_0\ln(1-n_3), \label{fi1}\\
\Phi^{(2)}&=&\frac{{\bf n}_1\cdot{\bf n_2}}{1-n_3}, \label{fi2}\\
\Phi^{(3)}&=&\frac{ n_{2x} n_{2y} n_{2z}}{(1-n_3)^2}, 
\label{fi3}
\end{eqnarray}
with weighted densities 
\begin{eqnarray}
n_{\alpha}({\bf r})=
\sum_{\mu}\left[\rho_{\mu}\ast \omega^{(\alpha)}_{\mu}\right]({\bf r}),
\label{weighted}
\end{eqnarray}
i.e., they are sums of convolutions of the density profiles with  
the following weights: 
\begin{eqnarray}
\omega^{(0)}_{\mu}({\bf r})&=&\frac{1}{8}
\prod_{k=1}^3 \delta\left(\frac{\sigma_{\mu}^{k}}{2}-|x_k|\right), \label{w0}\\
\omega^{(3)}_{\mu}({\bf r})&=&
\prod_{k=1}^3 \theta\left(\frac{\sigma_{\mu}^{k}}{2}-|x_k|\right), \label{w3}\\
{\omega}_{\mu}^{(1j)}({\bf r})&=&
\frac{2\theta\left(\frac{\sigma_{\mu}^{j}}{2}-|x_j|\right)}
{\delta\left(\frac{\sigma_{\mu}^{j}}{2}-|x_j|\right)}
\omega^{(0)}_{\mu}({\bf r}), \label{w1}\\
{\omega}_{\mu}^{(2j)}({\bf r})&=&
\frac{\delta\left(\frac{\sigma_{\mu}^{j}}{2}-|x_j|\right)}
{2\theta\left(\frac{\sigma_{\mu}^{j}}{2}-|x_j|\right)}
\omega^{(3)}_{\mu}({\bf r}), \label{w2}
\end{eqnarray}
where the notation 
$x_k$ ($k=1,2,3$) for the $x$, $y$, and $z$ coordinates has been employed. 
The functions 
$\delta(x)$ and $\theta(x)$ are the usual delta Dirac and Hevisaide functions   
and $\sigma_{\mu}^{j}=\sigma+(L-\sigma)\delta_{\mu}^j$ with 
$\delta_{\mu}^j$ the Kronecker delta. 

The following constraints on 
the density profiles were imposed:
(i) The solid phase 
has the simple parallelepipedic unit cell with uniaxial 
symmetry, i.e., 
the periods in the three spatial directions are $d_{\perp}$ for $x,y$ and 
$d_{\parallel}$ for $z$. The orientational director is selected 
parallel to $z$. 
 (ii) The density profile of each species has the form
\begin{eqnarray}
\rho_{\mu}({\bf r})=
\rho \gamma_{\mu}\sum_{{\bf k}={\bf 0}}^{{\bf n}}\alpha^{(\mu)}_{{\bf k}}
\prod_{j=1}^3\cos\left( q_jk_jx_j\right), \label{profile}
\end{eqnarray}
where $q_j=2\pi/d_j$ is the wave number along the $j$ direction, 
${\bf k}=(k_1,k_2,k_3)$ is 
the vector defined by the reciprocal lattice numbers, and 
${\bf n}=(n_1,n_2,n_3)$ is the vector at which the harmonic expansion
is truncated. 
Thus, Eq. (\ref{profile}) is the Fourier expansion of the density profiles 
$\rho_{\mu}({\bf r})$ truncated at some ${\bf n}$. This cutoff 
is selected in such a way that it guarantees small enough values of  
$\alpha_{{\bf n}}^{(\mu)}$.
The first Fourier 
amplitudes of all species are fixed to one 
($\alpha_{{\bf 0}}^{(\mu)}=1$) and consequently 
$V_{\rm{cell}}^{-1}
\int_{V_{\rm{cell}}} d{\bf r}\rho_{\mu}({\bf r})=\rho \gamma_{\mu}$ with 
$V_{\rm{cell}}=d_{\perp}^2d_{\parallel}$ the unit cell volume, 
$\rho$ the mean total density over the unit cell, and $\gamma_{\mu}$ 
the occupancy probability of species $\mu$ in the unit cell, which 
obviously fulfills 
the condition $\sum_{\mu}\gamma_{\mu}=1$.  

In the plastic solid phase these occupancy probabilities 
are $1/3$ for each species 
while they deviate from this value in the oriented solid phase.  
The uniaxial symmetry also implies that  
$\gamma_{x}=\gamma_{y}=\gamma_{\perp}$, $\gamma_{z}=\gamma_{\parallel}
=1-2\gamma_{\perp}$ and $\rho_x(x,y,z)=\rho_y(y,x,z)$, $\rho_z(x,y,z)=\rho_z(y,x,z)$. 
Thus, the Fourier amplitudes verify  $\alpha^{(x)}_{(k_1,k_2,k_3)}
=\alpha^{(y)}_{(k_2,k_1,k_3)}$ and 
$\alpha^{(z)}_{(k_1,k_2,k_3)}=\alpha^{(z)}_{(k_2,k_1,k_3)}$. 
The total number of Fourier amplitudes [except the $(0,0,0)$ term of 
all species] is reduced by these symmetries to 
$N_{\alpha}=(n_{\perp}+1)(n_{\parallel}+1)(3n_{\perp}+4)/2-2$, 
($n_1=n_2\equiv n_{\perp}$, $n_3\equiv n_{\parallel}$)
independent variables. 
These variables together with $\gamma_{\perp}$, $q_{\perp}$ and $q_{\parallel}$ span
the variable space in which the FMF must be minimized.

The density profiles of columnar and smectic phases are
obtained from Eq. (\ref{profile}) substituting 
${\bf n}=(n_{\perp},n_{\perp},0)$ and ${\bf n}=(0,0,n_{\parallel})$. 
From the definition (\ref{weighted}), Eqs. 
(\ref{w0})-(\ref{w2}) and the density profiles 
(\ref{profile}), the weighted densities can be 
easily calculated resulting in

\begin{eqnarray}
n_{\alpha}({\bf r})&=&\rho\sum_{\mu,{\bf k}}\gamma_{\mu}
\alpha^{(\mu)}_{{\bf k}}\chi_{\alpha,{\bf k}}^{(\mu)}\prod_{j=1}^3
\cos \left(q_jk_jx_j\right), 
\label{promediadas} \\
\chi_{0,{\bf k}}^{(\mu)}&=&\prod_{j=1}^3\phi_0\left(\xi_{j,{\bf k}}^{(\mu)}
\right), \\ 
\chi_{3,{\bf k}}^{(\mu)}&=&v\prod_{j=1}^3\phi_3\left(\xi_{j,{\bf k}}^{(\mu)}
\right), \\
\chi_{1j,{\bf k}}^{(\mu)}&=&\sigma_j^{\mu}\frac{\phi_3
\left(\xi_{j,{\bf k}}^{(\mu)}\right)}{\phi_0
\left(\xi_{j,{\bf k}}^{(\mu)}\right)}\chi_{0,{\bf k}}^{(\mu)}, \\
\chi_{2j,{\bf k}}^{(\mu)}&=&\frac{1}{\sigma_j^{\mu}}\frac{\phi_0
\left(\xi_{j,{\bf k}}^{(\mu)}\right)}{\phi_3
\left(\xi_{j,{\bf k}}^{(\mu)}\right)}\chi_{3,{\bf k}}^{(\mu)}, 
\end{eqnarray}
with $v=L\sigma^2$ the particle volume,  
$\phi_0(x)=\cos x$, $\phi_3(x)=\sin x/x$, and  
$\xi_{j,{\bf k}}^{(\mu)}=q_jk_j\sigma_j^{\mu}/2$.

The substitution of Eqs. (\ref{profile}) and  (\ref{promediadas}) 
into the free energy per unit cell
\begin{eqnarray}
\Phi\equiv\frac{\beta {\cal F}}{V_{\rm{cell}}}&=&V^{-1}_{\rm{cell}}
\int_{V_{\rm{cell}}} d{\bf r}
\left[\Phi_{\rm{id}}({\bf r})+\Phi_{\rm{exc}}({\bf r})\right], \label{perunit}\\
\Phi_{\rm{id}}({\bf r})&=&\sum_{\mu}\rho_{\mu}({\bf r})\left[
\ln \left(\rho_{\mu}({\bf r})\Lambda_{\mu}^3\right)-1\right],
\end{eqnarray}
with $\Phi_{\rm{id}}({\bf r})$ the ideal part of the free energy density, 
and its minimization with respect to the $N_{\alpha}+3$ variables 
allows the calculation of the equilibrium free energy and the density profiles
of inhomogeneous phases. 

To characterize the structure and orientational order 
of these phases  
the following total density and order parameter profiles will be used:
\begin{eqnarray}
\rho({\bf r})&=&\sum_{\mu}\rho_{\mu}({\bf r}), \label{total}\\
Q({\bf r})&=&1-\frac{3}{2}\frac{\left[\rho_x({\bf r})+\rho_y({\bf r})\right]}
{\rho({\bf r})}. \label{param}
\end{eqnarray}
The selection of $Q({\bf r})$ as an order parameter is motivated by its 
uniaxial symmetry property $Q(x,y,z)=Q(y,x,z)$ and its uniform limit value 
 $Q=1-3\gamma_{\perp}$ ($-1/2\leq Q\leq 1$), which coincides with the usual 
definition of the  nematic order parameter for the Zwanzig model: 
$Q=0$ ($\gamma_{\perp}=1/3$) for the isotropic phase and 
$Q=1$ ($\gamma_{\perp}=0$) for the perfectly aligned nematic phase. 
Although the solid and columnar phases
might have local biaxiality [$\rho_x(x,y,z)\neq \rho_y(x,y,z)$], 
the integral over the unit cell of any previously defined biaxial
order parameter is always equal to zero as a consequence of 
the symmetries of the density profiles.
\section{Phase diagrams}
\label{PD}
The phase diagrams presented in this work were calculated for a set of 
aspect ratios ranging from $\kappa=0.1$ to $\kappa=10$, corresponding 
to the aspect ratios of
the most anisotropic oblate and prolate parallelepipeds studied here. 
The volume of all particles (cubes or prolate or oblate parallelepipeds) 
are fixed to 1 and thus the mean packing fraction $\eta$ is equal to the 
mean density $\rho$. From the equation $v=L\sigma^2=1$ 
the parallelepiped edge lengths 
$L$ and $\sigma$ can be calculated as a function of the aspect ratio 
$\kappa=L/\sigma$ 
as $L=\kappa^{2/3}$ and $\sigma=\kappa^{-1/3}$.
For each $\kappa$, fixing the mean density $\rho$ 
and using appropriate initial guesses for 
the $N_{\alpha}+3$ variables with symmetries corresponding to the 
smectic, columnar or solid phases, the energy per unit cell (\ref{perunit}) 
was minimized and thus the free energy for each phase was obtained.
Varying $\rho$ and repeating 
the former steps the free-energy branches 
of the different inhomogeneous 
phases have been calculated. 
The common tangent construction allowed the calculation of the 
coexisting densities between those phases in the case of first order 
transitions. To evaluate numerically the three dimensional integral 
of the free energy density (\ref{perunit}) a  
Gauss-Chebyshev quadratures has been employed.  

\subsection{Parallel hard cubes}
\label{PHC}
This subsection is devoted to the study of the parallel hard cube 
system ($\kappa=1$). 
The PHC equation of states 
of the fluid and solid phases as obtained from the FMT and the Monte Carlo 
simulation results are compared. While the solid 
phase is very well described with this formalism the 
exact location of the fluid-solid transition is very poorly estimated. The
fundamental reasons of this difference are discussed here through a 
critical analysis of the fluid equation of state resulting from the FMF. 
It will be shown that possible modifications of the FMF slightly
improve the location of the transition point at the expense of 
the correct description of the solid branch. 

In Ref. \cite{Martinez-Raton_2} the PHC fluid was already studied with 
the same FMF but using a Gaussian parametrization for the 
density profile. Through a minimization procedure and 
also from a bifurcation analysis a second-order fluid-solid transition was found
at $\rho=0.3143$ with a lattice period $d=1.3015$ and 
a fraction of vacancies $\nu=0.3071$ \cite{Martinez-Raton_2}. 
Recent simulations 
on the same system also showed a second-order transition to the solid 
but with very different transition parameters 
$\rho=0.48\pm 0.02$ in Ref. 
\cite{Jagla} and $\rho=0.533\pm0.010$ in Ref. \cite{Groh}. 
No evidence for the vacancies predicted by FMT was found, although the authors 
recognized that the vacancies might be suppressed by the boundary 
conditions in the small systems accessible to simulations \cite{Groh}.

The main problem of the FMF for hard cubes is that it recovers in the 
homogeneous limit the scaled-particle equation of state, which overestimates 
the pressure calculated from the exact virial expansion up to seventh order. 
This expansion has a maximum at $\rho\approx 0.6$ and then goes down 
very quickly to reach negative values \cite{Swol}. The poorly convergent 
character of the virial series makes it impossible to construct an 
equation of state for hard cubes, such as the Carnahan-Starling equation 
for the hard-sphere fluid, which estimates reasonably well all the known virial
coefficients and diverges at close packing. On the other hand, it is well known that the 
FMF describes accurately the fluid structure in situations of high 
confinement, including the solid phase near close packing. For example, 
at high densities the functional recovers the cell theory, which is 
asymptotically exact when the packing fraction goes to 1, and also 
compares reasonably well with computer simulations \cite{Groh}. 
These nice properties are a consequence of a fundamental restriction, namely,
the dimensional cross-over \cite{Cuesta}, 
imposed in the construction of the FMF. The latter implies that the functional 
in dimension $D$ reduces to the functional in dimension $D-1$  
when the original density profile is constrained to $D-1$ dimensions, 
i.e., $\rho^{(D)}({\bf r})=\rho^{(D-1)}({\bf r})\delta(x_D)$, where $x_D$ is the 
coordinate that is eliminated on going from $D$ to $D-1$ dimensions.\\

One possible procedure to improve the description of the uniform fluid of 
hard cubes at the level of the FMF is to follow the same method used in Refs. 
\cite{Tarazona_2} and \cite{Roth}, in which the hard-sphere 
Carnahan-Starling equation of state is imposed through the modification
of the third term $\Phi^{(3)}$ [see Eq. (\ref{three})] of the excess 
free-energy density while keeping the exact density expansion of the 
direct correlation function up to first order. 
Unfortunately the absence of a good equation of state for the PHC fluid  
with the already mentioned properties makes this procedure less 
systematic compared to that of hard spheres \cite{Tarazona_2,Roth}.

Following this purpose the original excess free-energy density for hard cubes 
(\ref{three}) is now substituted by 
\begin{eqnarray}
\Phi_{\rm{exc}}({\bf r})=
\sum_{k=1}^3f_k\left(n_3({\bf r})\right)\Phi^{(k)}({\bf r}), \label{substitute}
\end{eqnarray}
with the $f_k(n_3)$'s selected in such a way as to keep the correct first order 
density expansion of the direct correlation function and to obtain the right 
virial expansion up to the seventh order of the PHC equation of state. As the 
original FMF for hard cubes gives the third virial coefficient correctly, these 
conditions imply that $f_{1,2}(n_3)\sim 1+O(n_3^2)$ and 
$f_3(n_3)\sim 1+O(n_3)$ for small $n_3$. Two further important 
conditions imposed on the $f_k(n_3)$'s are their limiting behavior when 
the local packing fraction tends to unity: $\lim_{n_3\to 1}f_{\alpha}
(n_3)=1$, which asymptotically guarantees the correct cell-theory limit, 
and the positive signs of their values, which guarantee the 
convexity of the fluid free energy. 
Unfortunately this procedure breaks the dimensional cross-over 
property, but in principle should describe the 
fluid-solid transition in hard cubes better.

\begin{figure}
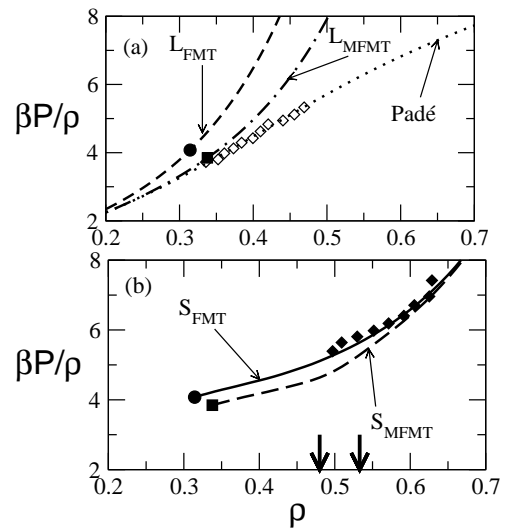

\mbox{\includegraphics*[width=2.5in, angle=0]{fig1a.eps}}
\hspace*{0.5mm}
\mbox{\includegraphics*[width=2.56in, angle=0]{fig1b.eps}}
\caption{(a) Equations of state of the PHC liquid 
following the FMT ($L_{\rm{FMT}}$), the modified version  
($L_{\rm{MFMT}}$), and the symmetric Pad\'e approximant.  
The circle and square represent the location of bifurcation 
points of the fluid-solid transition. 
(b) The equations 
of state of the solid phase from the original FMT ($S_{\rm{FMT}}$) and from 
the modified version ($S_{\rm{MFMT}}$). The arrows represent the fluid-solid
transitions predicted in Ref. \cite{Jagla} (the lower value) 
and Ref. \cite{Groh} (the higher value). 
Open and black 
diamonds are the simulation results from Ref. \cite{Jagla} 
corresponding to the liquid and solid phases, respectively.}
\label{fig1}
\end{figure}

Among all the functions $f_{\alpha}$'s that have been tried, even those  
which give better results [the particular case of 
$f_{1,2}(n_3)=1$] are far from getting 
the transition point near the simulation one. In Fig. 
\ref{fig1}(a) the scaled-particle equation of state, the improved 
equation of state 
\begin{eqnarray}
\beta P=\rho+\rho \frac{\partial \Phi_{\rm{exc}}}{\partial \rho}-
\Phi_{\rm{exc}},
\end{eqnarray}
with $\Phi_{\rm{exc}}$ being the uniform limit of 
Eq. (\ref{substitute}), and finally, 
the symmetric Pad\'e approximant of the seventh-order virial series are plotted.
In the first two curves the bifurcation points are shown. The new 
bifurcation point calculated from Eq. (\ref{substitute}) is located at 
$\rho=0.3378$, and the period and fraction of vacancies of the solid are 
$d=1.3249$ and $\nu=0.2143$. As can be seen from Fig. \ref{fig1}(a),
the new equation of state still overestimates the fluid pressure, 
but to a lesser 
extent. Although the new functional gets a higher transition density and the 
fraction of vacancies decreases, there is still 
disagreement between theory and simulations. The equation 
of state of the PHC solid calculated from the minimization 
of the original FMF with respect to the Gaussian density profiles 
compare very well with simulations
for densities $\rho\gtrsim 0.5$, whereas 
the modified version underestimates the solid pressure.

At this point the main conclusion that can be drawn is that the 
modification of the FMF in order to improve the description of 
the uniform fluid spoils the good description of the solid phase.
As the modification of the FMF 
was done at the expense of loosing the dimensional cross-over 
property (and this spoils the good description of highly inhomogeneous 
phases), and the modified versions do not show too many  
differences in the prediction of the fluid-solid transition, 
it is worthless to use them to study nonuniform phases.   

\begin{figure}
\mbox{\includegraphics*[width=2.5in, angle=0]{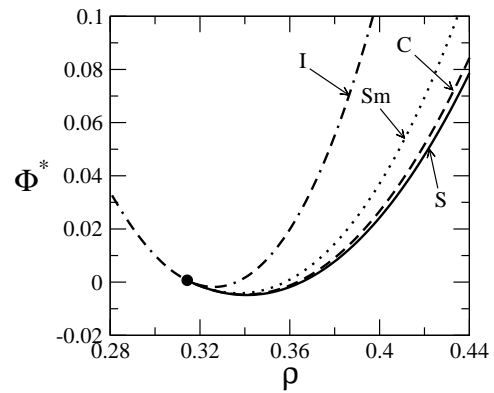}}
\caption{Free energy per unit cell as a function of the mean density $\rho$. 
A linear function of $\rho$ was subtracted from the free energy 
$\Phi^{\ast}\equiv\Phi-m\rho-n$ to make clear the energy differences between 
the isotropic ($I$), columnar ($C$) and solid ($S$) phases.} 
\label{fig2}
\end{figure}
Setting $q_{\parallel}=q_{\perp}=q=2\pi/d$, $\gamma_{\mu}=1/3$, and 
$\alpha_{\bf k}^{(\mu)}=\beta_{\bf k}$ in the
density profiles (\ref{profile}) and minimizing 
the FMF, Eq. (\ref{perunit}), of parallel hard cubes ($\kappa=1$)  
with respect to the Fourier amplitudes and the wave number $q$, 
the free energy per unit cell for solid 
[${\bf n}=n(1,1,1)$], columnar [${\bf n}=n(1,1,0)$] and 
smectic [${\bf n}=n(0,0,1)$] phases were obtained. 
The results are shown in Fig. \ref{fig2}. 
From the isotropic liquid at the same density $\rho=0.3134$ three 
inhomogeneous solutions: solid, columnar, and smectic, bifurcate,
with the solid phase being the stable one.   
While the free energy difference between solid and columnar 
phases is relatively small, the smectic phase is clearly thermodynamically
unfavorable. 

The number of Fourier amplitudes 
necessary to describe adequately the density profile increases with 
the density, and thus the numerical calculations becomes more and more 
time consuming. Nevertheless, the scenario shown in Fig. \ref{fig2}, 
with the solid being the only stable phase, occurs at high densities
as the simulations and cell-theory have confirmed
\cite{Groh}. The minimization of the FMF using a Gaussian 
parametrization of the density profiles of columnar and solid phases 
shows very similar quantitative results \cite{Groh}.  
In fact the equation of state of the parallel hard-cube 
solid from FMT calculations with this parametrization 
compares very well at high densities with simulations \cite{Martinez-Raton_3}. 
The results  presented here are much more accurate than those obtained 
through the Gaussian parametrization. 
\subsection{Prolate parallelepipeds}
\label{PP}
\begin{figure}
\mbox{\includegraphics*[width=2.5in, angle=0]{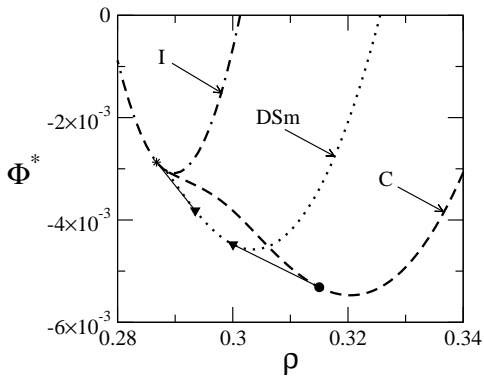}}
\caption{Free energy per unit volume as a function of the mean density $\rho$. 
The involved phases are labeled as in Fig. \ref{fig2}. 
DSm: discotic smectic phase. The common-tangent constructions, 
which determine the coexisting densities labeled with different symbols,
are also shown.} 
\label{fig3}
\end{figure}

\begin{figure}
\mbox{\includegraphics*[width=2.5in, angle=0]{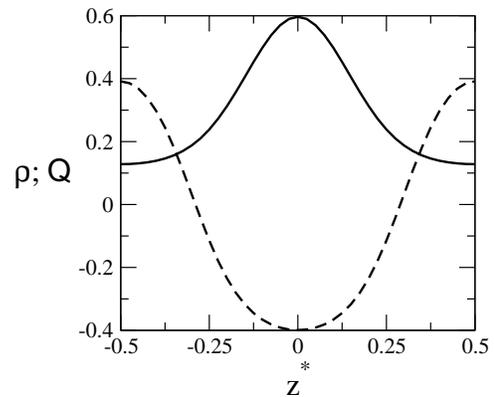}}
\caption{Density profile (solid line) and order 
parameter profile (dashed line) of the DSm phase at $\rho=0.3$ 
($z^*\equiv z/d$).} 
\label{fig4}
\end{figure}

This subsection is devoted to study the phase diagram of prolate 
particles ($\kappa>1$).
The results obtained from numerical minimization of the FMF of parallelepipeds 
with fixed $\kappa=4.5$ are shown in Fig. \ref{fig3}. 
The free energies per unit volume of those phases 
which are stable in some range 
of densities are plotted. As can be seen the isotropic phase undergoes a first-order 
phase transition to the so-called discotic smectic (DSm) phase. This peculiar 
phase is a layered phase (similar to the smectic phase) 
but with the long axes of the 
parallelepipeds lying within the layers. There is no orientational order in
the layers, what means that the order parameter $Q(z)$ reaches negative 
values at the positions of the density peaks. The density and 
order parameter profiles of the DSm phase at $\rho=0.3$ 
are plotted in Fig. \ref{fig4}. 
The period in units of the small 
particle length is $d/\sigma=1.2796$ which means that the particles 
with long axes perpendicular to the layers (preferentially localized at 
the center of the interlayer space) intersect 
approximately three adjacent layers. 

Simulations of the Zwanzig model with $\kappa=5$ on a lattice showed 
an I-DSm transition at a density between 0.47 and 0.55 \cite{Casey}. 
Although the results were obtained for a lattice spacing of $1/3$ (in units 
of the shortest particle dimension) the simulations were repeated for 
values $1/9$ and $1/27$ without changes in the stability 
of the DSm phase. Thus, the authors concluded that this layered 
phase may persist in the continuum limit \cite{Casey}. The difference 
in the transition density found from FMT ($0.2868$) and from simulations 
($\sim 0.5$) can be explained using two arguments: 
(i) As was already pointed 
out in Sec. \ref{FMF}, the FMF in the uniform density limit 
considerably overestimates the isotropic fluid pressure and thus the theory 
underestimate the transition densities  between homogeneous and inhomogeneous phases.  
(ii) The transition densities 
should decrease upon decreasing the lattice spacing in simulations, 
as the results for the freezing of parallel hard cubes 
on a lattice (occurring at $\rho=0.568$ for an edge length equal to two lattice 
spacings, at $\rho=0.402$, for six lattice spacings, and at $\rho=0.314$ for the 
continuum) illustrate \cite{Lafuente}.

Increasing the mean density further the DSm phase undergoes a first-order  
transition to the columnar phase as Fig. \ref{fig3} shows. 
The restriction of parallelepiped orientations  
enhances the columnar phase stability even for prolate 
particles as a phase diagram, to be described below, will show. 
This phenomenon can be understood if the Zwanzig model is interpreted as 
a ternary mixture of particles. Simulations on a binary mixture of parallel 
spherocylinders with different aspect ratios (specifically 2 and 2.9) 
show that, instead of a continuous nematic-smectic transition typical of the 
pure component system, the mixture exhibits a first-order nematic-columnar 
phase transition \cite{Stroobants_2}. 
This result was explained by the poorer packing 
of rods of different lengths in the smectic phase as compared to that
of rods of the same length. Simulations and theory show that 
one of the most important effects that the aspect ratio polydispersity 
has on the phase behavior of hard spherocylinders \cite{Polson} and 
binary mixtures of oblate and prolate particles \cite{vanKoij,
Martinez-Raton_3} 
is the enhancement of the columnar phase stability. All these results 
show that the columnar phase can be stable even for mixture of particles with 
different shapes. Although the constituent particles of the Zwanzig 
model have the same shape, the restriction of their orientations changes 
strongly its relative packing and thus for some $\kappa$'s 
enhance the columnar phase stability with respect to other phases.

\begin{figure}
\mbox{\includegraphics*[width=2.5in, angle=0]{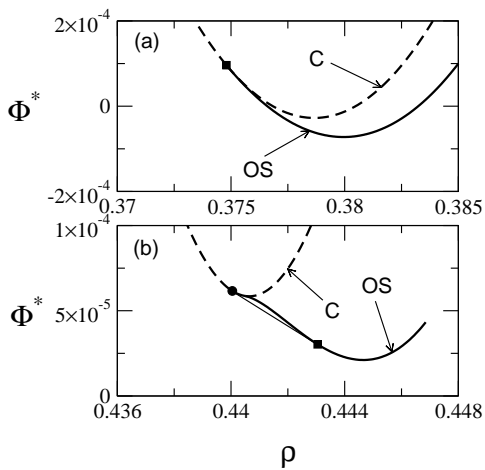}}
\caption{ $\Phi^*$ vs $\rho$ for $\kappa=4.5$ (a) and $\kappa=4$ (b). 
OS: oriented solid phase.} 
\label{fig5}
\end{figure}

\begin{figure}
\mbox{\includegraphics*[width=3.2in, angle=0]{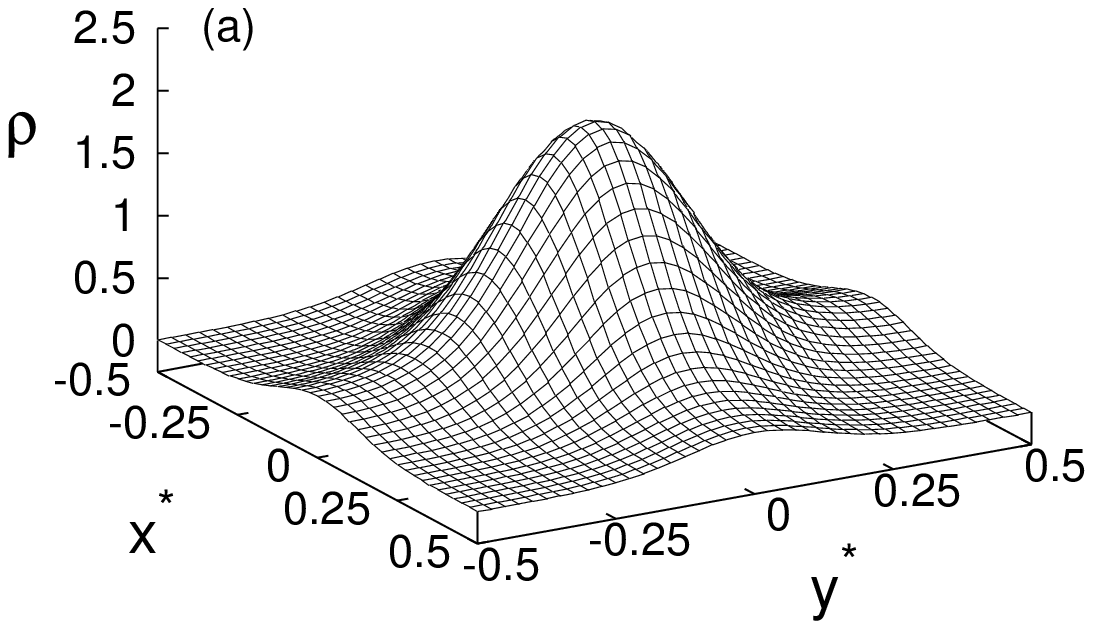}}
\mbox{\includegraphics*[width=3.2in, angle=0]{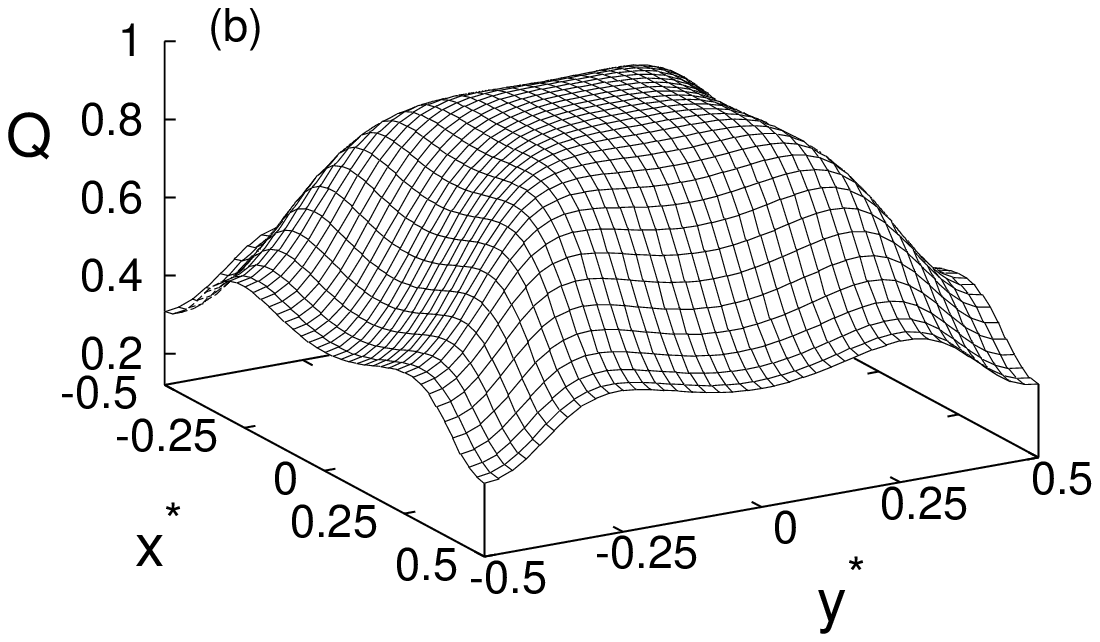}}
\caption{ Density (a) and order parameter (b) profiles of the columnar 
phase at a density corresponding to the bifurcation point of 
Fig. \ref{fig5}(a) ($x^*\equiv x/d_{\perp}$; $y^*\equiv y/d_{\perp}$).} 
\label{fig6}
\end{figure}

\begin{figure}
\mbox{\includegraphics*[width=3.2in, angle=0]{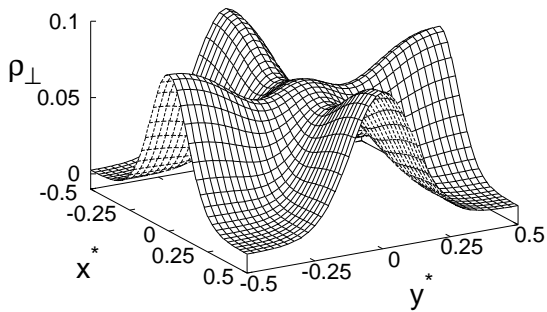}}
\caption{Sum of density profiles: 
$\rho_{\perp}({\bf r})\equiv\rho_x({\bf r})+\rho_y({\bf r})$ 
vs $r_{\perp}^*\equiv(x^*,y^*)$ 
corresponding to the columnar phase shown in Fig. 
\ref{fig6}.}
\label{fig7}
\end{figure}

\begin{figure}
\mbox{\includegraphics*[width=2.5in, angle=0]{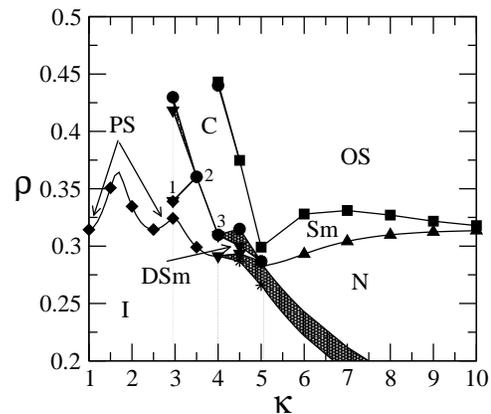}}
\caption{Phase diagram of prolate parallelepipeds. Several phases are 
labeled like in Figs. \ref{fig2}, \ref{fig3}, and \ref{fig5} ($N$: nematic and 
$PS$: plastic solid) and the transition densities are labeled with 
different symbols (Star: isotropic; diamond: plastic solid; circle: columnar; 
down triangle: discotic smectic; up triangle: smectic; and square: oriented 
solid). The shaded areas limit the regions of two phase coexistence. 
The transitions labeled by 
by 1,2, and 3 are first order in nature.} 
\label{fig8}
\end{figure}

At higher density the columnar phase exhibits a continuous phase transition 
to an oriented solid phase of prolate parallelepipeds, as shown in 
Fig. \ref{fig5}(a). The density and order parameter profiles of the 
columnar phase at the bifurcation point  ($\rho=0.3748$) are 
shown in Fig. \ref{fig6}. The periods of the solid phase 
along the perpendicular and parallel directions
are $d_{\perp}/\sigma=1.2690$ and 
$d_{\parallel}/L=1.5170$, respectively. From the equation $\rho=(1-\nu)
V_{\rm{cell}}^{-1}$ ($V_{\rm{cell}}=d_{\perp}^2d_{\parallel}$ being the 
unit cell volume), the fraction of vacancies of the solid $\nu$
can be calculated as $0.0845$. 
The continuous nature of the columnar-oriented solid transition changes to 
first order at some $\kappa$ between $4$ and $4.5$, as Fig. 
\ref{fig5}(b) shows for $\kappa=4$.
The order parameter $Q({\bf r})$ 
is very high in the unit cell except in its borders, 
where it exhibits small oscillations [see Fig. \ref{fig6}(b)]. 
These oscillations are a consequence of the 
microsegregation of species ``$x$'' and ``$y$'' in the newly formed 
solid phase, which is preferentially formed by particles of
species ``$z$'' localized 
around the position $(x^*,y^*)=(0,0)$. 
This feature is shown in Fig. \ref{fig7}, where the sum of the density
profiles of species ``$x$'' and ``$y$'' 
[$\rho_{\perp}({\bf r})=\rho_x({\bf r})+
\rho_y({\bf r})$] is plotted. While the columnar packing is responsible 
for the presence of the local maxima  at the center of the unit cell, 
the species ``$x$'' and ``$y$'' begin to segregate to the borders of 
the cell $(\pm 0.5,0)$ and $(0,\pm 0.5)$, respectively (see the four 
local maxima at these positions) as the new solid phase is formed. 
The long axes of the perpendicular species lie on the lateral  
sides of the parallelepipedic unit cell, while their centers of 
mass are preferentially localized at the centers of these sides.\\ 
The calculation of the free-energy branches for several stable inhomogeneous phases and 
the phase transitions between them (as it was described for $\kappa=4.5$) 
has been carried out for 15 values of 
$\kappa$ (ten of them in the range $1\le\kappa\le 5$ and 
five of them in the range $5\le\kappa\le 10$). The resulting 
phase diagram is plotted in Fig. \ref{fig8}. 
The isotropic phase of prolate parallelepipeds 
with $1\leq \kappa \leq 3.5$ undergoes a continuous 
phase transition to the plastic solid phase. The transition 
points are joined with the spinodal line that has been calculated through 
the divergence of the structure factor. Notwithstanding that a functional 
minimization was carried out for each $\kappa$ to check the continuous 
nature of the transitions.  The plastic solid is stable for 
$\kappa\leq 2.5$ up to densities around $0.5$. At these 
values the numerical minimization turns out to be cumbersome because 
of the large number of Fourier amplitudes necessary to correctly describe the 
inhomogeneous profiles. Thus, the high density part of the phase diagram 
($\rho \gtrsim 0.5$) has not been calculated with 
the numerical procedure described above. 
At higher densities a Gaussian-type parametrization 
of the density profiles is required, which obviously 
has a lower degree of accuracy.  

For $\kappa=2.95$ the plastic solid exhibits a very weak 
first-order phase transition 
to the discotic smectic phase (labeled as 1 in Fig. \ref{fig8}), and the 
latter a phase transition to the columnar phase at higher densities. 
But the most representative 
region of the phase diagram where the discotic smectic is stable is 
for $\kappa$ around 4.5 where this layered phase exhibits a first-order 
phase transition to columnar phase (the shaded area of Fig. \ref{fig8} limits 
the instability region against phase separation between both phases). 
For $\kappa$ between 4 and 5 the columnar phase undergoes a phase transition 
(first order for $\kappa=4$ and continuous for other values shown) to the oriented 
solid phase. 
The nematic phase begins to be stable 
for $\kappa >5$ with its stability region bounded 
below by the first order isotropic-nematic 
transition and above by a continuous nematic-smectic transition. Finally, 
the smectic region is bounded above by a continuous transition to 
the oriented solid phase (see Fig. \ref{fig8}).  

Again the nematic-smectic transition points are joined with 
spinodal lines and for each $\kappa$ 
a minimization was carried out to check numerically 
the continuous character of 
the transition (the smectic solution begins to be stable right at 
the spinodal). 
In Ref. \cite{Martinez-Raton_5}
a bifurcation analysis with the same 
functional was carried out to study 
the nature of the nematic-smectic transition. 
A thermodynamic and mechanical 
stability analysis showed that the 
nematic-smectic transition is  first order,
which is in contradiction with the numerical minimization results 
presented here. 
A possible reason that justifies this contradiction could be that
the $N$-Sm transition is very weakly first order,   
so weak that the numerical accuracy used in the functional 
minimization can not decide about its nature. Another possibility is 
that the numerical accuracy failure is somewhere in the bifurcation 
analysis. A careful revision of this analysis is certainly called for 
in order to settle this point. 

The available simulation results for freely rotating hard spherocylinders 
show that the isotropic phase exhibits a transition to the solid phase 
for $0\leq\kappa\leq 4.1$ (the solid is plastic for $\kappa\leq 1.35$ and 
oriented for $1.35\leq \kappa\leq 4.1$) while the
isotropic-smectic and nematic-smectic transitions begin at 
$\kappa=4.1$ and $4.7$, respectively \cite{Bolhuis} [notice that for hard
spherocylinders the length-to-breadth ratio is $\kappa=(L+D)/D$].
We can see that, despite the different particle geometry and the restricted 
orientations of the Zwanzig model, the agreement for the 
threshold $\kappa$ at which spatial instabilities to the solid 
and smectic phase destabilize the homogeneous phases is rather 
good. Also the qualitative picture is similar: elongated rods form smectics, 
and more symmetric particles form solids. 
The main difference between them is that the Zwanzig phase diagram 
presents regions where the columnar and discotic smectic 
phases are stable, a 
difference due to the restriction of orientations.  
\subsection{Oblate parallelepipeds}
\label{OP}
\begin{figure}
\mbox{\includegraphics*[width=2.5in, angle=0]{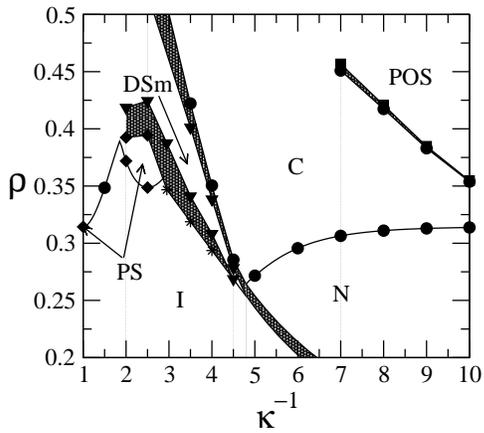}}
\caption{Phase diagram of oblate parallelepipeds
(the same symbols and labels of Fig. \ref{fig8} are used). 
POS: Perfectly oriented solid.} 
\label{fig9}
\end{figure}

The phase diagram of oblate parallelepipeds ($\kappa<1$) is shown in Fig. 
\ref{fig9}. The main differences after comparing the phase diagrams of 
prolate  (Fig. \ref{fig8}) and oblate particles are that in the latter:  
(i) The smectic is no more a stable phase. (ii) The region 
of columnar phase stability is considerably larger. (iii) The  
stability region of the plastic solid is reduced (in fact this phase is 
stable only up to $\kappa^{-1}\approx 2.5$) at the expense of that 
of the discotic smectic phase.  
(iv) The transitions to the 
latter are strongly first order in nature (except for $\kappa^{-1}=4.5$).
(v) The oriented solid 
phase is replaced by a perfectly oriented solid in which ``$x$'' and ``$y$'' 
species are absent. This phase, after scaling in the $z$ direction, 
is the same as the solid of parallel hard cubes.  
A solution from the FMF minimization with three dimensional 
spatial modulations and with 
$\gamma_{\perp}\neq 0$ has not been found in the 
parallelepipedic unit cell
(the case of face-centered or body-centered cubic unit cells have 
not been tried here).  

Finally, similar by to what happens with prolate parallelepipeds, the 
nematic phase begins to be stable at $\kappa^{-1} \gtrsim 5$. 
It undergoes a continuous phase 
transition to the columnar phase (the transition points of 
Fig. \ref{fig9} are joined with the spinodal curve).

The parallelepipeds with $\kappa^{-1}=1.5$ exhibit an interesting 
phase behavior. At low densities the isotropic phase destabilizes with 
respect to the columnar phase and not with respect to the PS phase. 
This example shows that the prediction for phase transitions 
using only the spinodal instability calculations can generate uncertainties  
about the possible symmetries of the inhomogeneous phases. In fact these 
calculations do not allow to decide in this example 
if the new phase is a plastic solid or a columnar phase. 
Only by a complete minimization of the FMF, could it be concluded 
that the columnar 
phase is the stable one.

\begin{figure}
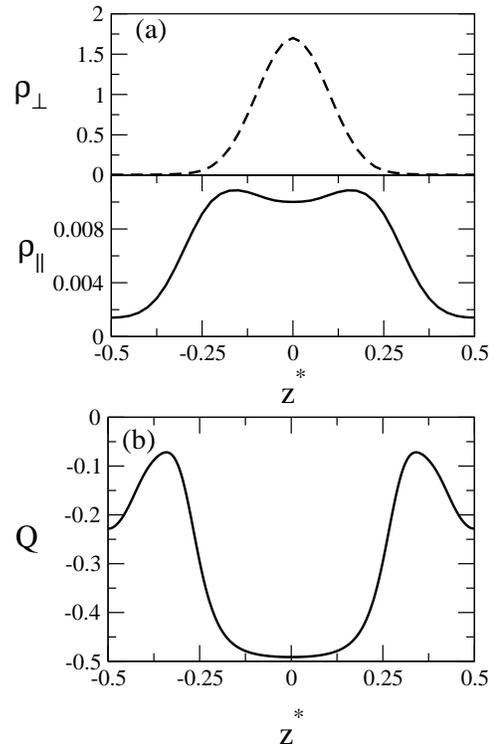

\mbox{\includegraphics*[width=2.5in, angle=0]{fig10a.eps}}
\mbox{\includegraphics*[width=2.5in, angle=0]{fig10b.eps}}
\caption{Density (a) and order parameter (b) profiles of the DSm phase 
coexisting at $\rho=0.4244$ with the PS. (a) Shows the 
density profiles of perpendicular ($\rho_{\perp}$) and parallel 
($\rho_{\parallel}$) species.}
\label{fig10}
\end{figure}

In Fig. \ref{fig10} (a) the density profiles of perpendicular 
[$\rho_{\perp}(z)$] and parallel [$\rho_{\parallel}(z)$] species  
are shown for the discotic smectic phase of oblate particles with $\kappa^{-1}=2.5$,
while the order parameter profile is shown in Fig. \ref{fig10}(b). 
This discotic smectic phase coexists at $\rho=0.4244$ with the plastic solid phase. 
The period in units of the large parallelepiped edge length is $d/\sigma=1.2142$. The 
random orientation of the uniaxial axes within the layers is confirmed 
by the high negative values of the order parameter at the 
position of the density peak of the perpendicular species. 
The main difference between the DSm of 
Fig. \ref{fig10} and that of Fig. \ref{fig7} is that the ``z'' species 
is now localized preferentially not at the center of the interlayer space, 
but near the smectic layers [see Fig. \ref{fig10}(a)], exhibiting two local 
maxima at each side of the layer. This effect can 
be explained by the depletion force that the perpendicular species exerts
on the parallel one.\\ 
The $N$-Sm ($N-C$) and the Sm-OS ($C$-POS) transition lines 
of Figs. \ref{fig8} and  
\ref{fig9} converge asymptotically 
to $\rho=0.3143$, the value of the fluid-solid bifurcation 
density, as $\kappa\to \infty$ ($\kappa^{-1}\to \infty$). The reason 
for this is that upon increasing $\kappa$ ($\kappa^{-1}$) the number of rods 
(plates) with orientation perpendicular to the director 
becomes vanishing small, 
and then the system is, after rescaling the $z$ direction, almost equivalent 
to a system of parallel cubes. 

Simulations of the Zwanzig model on a $15\times 15\times 15$ 
lattice with spacing $1/3$ show that oblate 
parallelepipeds with dimensions $5\times 5\times 1$ undergo a transition 
to a phase exhibiting a columnar structure \cite{Casey} 
at a density somewhere between $0.55<\rho<0.65$. 
On increasing the system size to $30\times 30\times 30$ 
the global columnar order disappears, but local correlations persist in the 
fluid with particle alignment distributed evenly among the three available 
orientations. In the same work the effect that orientational 
constraints have on the stability of the inhomogeneous phases was studied. 
While a system 
of biaxial $5\times 3\times 1$ ``tiles'' without orientational 
constraints (except those 
inherent to the Zwanzig model) stabilizes in a smectic phase with the shortest 
edge lengths perpendicular to the layers, the system composed by ``tiles'' with 
all their long edge lengths parallel to each other 
exhibits a phase transition to 
the smectic phase with these edge lengths perpendicular to the layers, 
similar to what is found here for uniaxial oblate parallelepipeds (the 
discotic smectic phase).  

Simulations of hard cut spheres show that for $\kappa=0.3$ there is 
an isotropic-solid transition, for $\kappa=0.2$ an 
isotropic-columnar transition (the isotropic phase might instead be a 
peculiar ``cubatic'' phase) and for $\kappa=0.1$  
a nematic-columnar one \cite{Veerman}. From these results it can be 
concluded that the effect that the 
degree of particle anisotropy has on the symmetry of the stable  
phases for both cut spheres and hard parallelepipeds with 
restricted orientations, is qualitatively similar. 
\section{Conclusions}
The goal of this article has been the calculation of the phase diagram of the 
Zwanzig model for prolate and oblate parallelepipeds centering the 
attention on the phase transitions to inhomogeneous phases. For this purpose 
the fundamental measure functional for hard parallelepipeds 
with restricted orientations has been used. 
This functional is exactly the same for any 
particle shape (prolate and oblate depending on $\kappa$), which allows for a
unified study of the phase behavior of both kinds of particles. 
A free minimization of the functional was carried out
with the only constraints of choosing a  
parallelepipedic unit cell and of imposing uniaxial symmetry 
in the inhomogeneous phases. 
The latter is justified by uniaxial 
symmetry of the particles. The degree of approximation 
to the exact density profiles was 
controlled by the cutoff imposed on the reciprocal lattice numbers  
in the Fourier expansion of the density profiles.

A system of parallel hard cubes was separately studied, which was motivated by 
recent simulations on this system \cite{Jagla,Groh}. Applying a modified 
versions of the FMF to improve the description of the PHC liquid, 
along the same lines as Refs. \cite{Tarazona_2} and \cite{Roth}, 
the continuous transition point to the solid phase and the  
equation of state of the solid 
were calculated from the divergence of the 
structure factor and from the functional minimization 
with respect to  
Gaussian density profiles.  
Although the transition density and fraction of vacancies change 
in the right direction, these results are still far from the simulations. 
In fact, the solid phase is poorly described by the new functional.
The poor convergence of the PHC virial 
series does not make this procedure as effective as for hard spheres. 
Further refinement of the method and the proper 
inclusion of vacancies in simulations of the solid phase 
will probably improve the agreement between theory and simulations. 

The original FMF for PHC was minimized 
to study the relative stability of the smectic, columnar, and solid phases,
starting at low densities from the bifurcation point. The solid phase is the only 
stable phase, followed by the columnar and the smectic 
(in order of energy stability). At high densities the same behavior is shown  
from calculations using cell theory, functional minimization 
with Gaussian density profiles and computer simulations \cite{Groh}.

The system of prolate and oblate parallelepipeds exhibits a very rich phase 
behavior. Apart from the plastic or oriented solid, smectic, and columnar 
phases, which are present also in systems of prolate (spherocylinders 
\cite{Bolhuis}) and oblate (cut spheres \cite{Veerman}) 
particles, a new phase appears: the discotic smectic, the existence of 
which was confirmed by simulations \cite{Casey}. The close 
relation between the particle anisotropy and symmetry of the stable phases 
(elongated particles form smectics, flattened one form columnars and more 
isotropic particles form solids) which has been observed in simulations 
\cite{Strobants_1, Bolhuis, Veerman} and experiments \cite{Maeda} 
is confirmed by this simple model. 

There are two 
important effects that the restriction of orientations has on 
the phase diagram topology: (i) The already pointed out stability 
of the discotic smectic. (ii) The stability of the columnar 
phase of prolate parallelepipeds for some aspect ratios. 
The structural properties of inhomogeneous phases that were found through 
functional minimization allow us to elucidate interesting effects such as
the microsegregation behavior of different species in the solids and 
the depletion effect between particles in the smectics. Those findings 
endorse the predictive power of the FMF in the description of highly 
inhomogeneous phases. 

\begin{acknowledgments}
The author thanks J. A. Cuesta and E. Velasco for 
useful discussions and 
a critical reading of the manuscript and E. Jagla for kindly providing 
his simulations data. This work is part of the research 
Project No. BFM2004-0180 (DGI) of the 
Ministerio de Ciencia y Tecnolog\'{\i}a (Spain). The author was 
supported  by a Ram\'on y Cajal research contract.
\end{acknowledgments}

\end{document}